\documentclass[pra,aps, english, twocolumn, hyperref]{revtex4}

\usepackage{graphicx}

\usepackage{amssymb, amsmath, color, rotating}
\begin{document}

\title{Dynamics of correlations in a dilute Bose gas following an interaction quench}

\author{Stefan S. Natu}

\email{ssn8@cornell.edu}

\affiliation{Laboratory of Atomic and Solid State Physics, Cornell University, Ithaca, New York 14853, USA.}

\author{Erich J. Mueller}

\affiliation{Laboratory of Atomic and Solid State Physics, Cornell University, Ithaca, New York 14853, USA.}

\begin{abstract}
We calculate the dynamics of one and two body correlation functions in a homogeneous Bose gas at zero temperature following a sudden change in the interaction strength, in the continuum and in a lattice. By choosing suitable examples, we highlight features in the correlation functions that emerge due to the interactions and the band structure. We find that interactions dramatically change the way correlations build up and subsequently decay following a quench. For example, the Bogoliubov dispersion induces a crossover from diffusive spreading of short range correlations to ballistic spreading of longer range correlations. In the lattice, the correlation functions develop additional features absent in the continuum. Most strikingly, the lattice induces an additional velocity scale and some features propagate with that velocity. Finally, we discuss the ultra short-range properties of the density-density correlation function following a quench, and the implications for experiments using this quantity to probe the ``contact". Our calculations, which can be readily tested in current experiments, suggest that the dynamics of correlations may be a useful tool for extracting many-body parameters.

%In the continuum, we show that the Bogoliubov spectrum  leads to a diffusive evolution of short-range density correlations, while longer range correlations spread ballistically. By contrast, in the non-interacting system, no such crossover is found. Interactions also dramatically alter the very long time, asymptotic behavior of the correlation functions following the quench. %Once correlations are built up after the quench, they decay \textit{exponentially} in the interacting gas, as opposed to \textit{algebraically} in the non-interacting system. 
%In the lattice we show that the correlation functions develop additional features, absent in the continuum. For example, the bounded lattice spectrum induces an additional velocity scale, and some features instead propagate with that velocity. Finally, we discuss the time-evolution of the contact following a quench.  Our calculations point to the richness of non-equilibrium dynamics even for \textit{weak} interactions, and serve as a benchmark for understanding the properties of correlations in strongly interacting systems. 
\end{abstract}
\maketitle

\section{Introduction} 
New higher resolution imaging techniques are allowing cold atom experiments to probe spatial correlation functions \cite{chengstruc, greiner}.  Recently, the focus has turned to the dynamics of these correlations following a sudden change in experimental parameters such as lattice or trap depth or interaction strength \cite{blochlightcone, trotzky, demarco, miyakebragg, greinercollapse, chengnew, imambekov}.  In many ways, experiments are ahead of theory, as the very paradigms for thinking of these highly nonequilibrium experiments are just being developed \cite{cardy, kehrein, kollathnoneq}.  Here we use the time dependent Bogoliubov approximation to study quenches in a weakly interacting Bose gas.  We are  thereby able to organize the phenomena, for example separating out interaction effects from lattice effects.  Although experiments to date have been in a strong or intermediate coupling regime and cannot be directly modeled by our technique, the existing data appears to be consistent with these organizational principles.

%There are a number of fundamental issues which can be probed through the time dependence of correlation functions.  For example, experiments on lattice bosons have been related directly to questions of causality \cite{blochlightcone, liebrobinson, cardy}.  A second interesting direction is to make connections to 
%the fluctuation theorems in classical statistical mechanics \cite{jarzynski} which relate equilibrium free energy differences to the work done.  
%This latter physics has been studied in mesoscopic solid state systems by looking at the counting statistics of electrons tunneling through quantum dots \cite{fcs2}.    A third direction is using these quenches to help understand how isolated quantum systems approach equilibrium, and determine the timescale for the development of long and short range order \cite{schmiedmayerrel, sautebd, kehrein, kollathnoneq, weiss}. Our focus is in a fourth direction, namely on how quenches reveal the structure of the excitation spectrum. 

There are a number of fundamental issues which can be probed through the time dependence of correlation functions.  For example, studies of lattice models have found that the manner in which correlations develop following a quench is directly related to questions of causality \cite{blochlightcone, liebrobinson, cardy, natuexp, kollathferm, salvatoremtm, rigolmtm, polkovnikov}. This has important implications for understanding how quantum systems approach equilibrium and the timescales involved. Another key issue probed by these studies is the nature of the final state obtained at long times following such a quench, and the degree of short or long range order \cite{schmiedmayerrel, sautebd, kehrein, kollathnoneq, weiss, kibble, fischer}.  Here we explore how the underlying dispersion influences the spatio-temporal characteristics of the correlation functions.

Typically the time-evolution of correlation functions can be quite complicated \cite{blochlightcone, kollathferm, trotzky} and non-intuitive. Our aim in this paper is to organize the salient features in the density-density correlation function for a weakly interacting Bose gas, whose static properties are textbook knowledge for atomic physicists \cite{pethick, stringari, glauber, lhy}. Using time dependent Bogoliubov theory, we calculate how the correlation functions evolve following a sudden change in the interaction strength.  We find that in free space the density correlations spread diffusively at short times, crossing over to ballistic at long times. The speed of propagation is the phonon velocity. Furthermore, the decay of the correlation functions at long times is strongly influenced by the underlying dispersion. 

As many of the strongly correlated models are lattice models, separating the role of the lattice from that of the interactions is crucial.  We find that in the lattice, new features appear which propagate with a speed determined by the bandwidth.  The speed of propagation crosses over to the sound speed at long times.  %We also look at single particle correlations, for example finding that the time for the condensate fraction to reach equilibrium is simply related to  both the initial and final chemical potential.  

In addition to ``long wavelength" physics, we find that the short length-scale behavior of the correlation functions has interesting structure. Immediately following the quench, the two-particle correlation function develops a divergence which scales as $1/|\textbf{r} - \textbf{r}^{'}|^{2}$. This structure is due to the singular nature of the two-particle relative wave function at short distances \cite{lhy}. In equilibrium, this singularity has attracted a lot of attention recently following the work of Tan, who was able to relate the short distance structure of the two-particle correlation function in a two-component Fermi gas to the internal energy via a quantity called the ``contact" \cite{tan, vale, stoof, jincontact}. Here we study the dynamics of the contact (and its generalizations) following a quench.   

This paper is organized as follows: In Sec. II we describe our system, derive in detail the equations of motion governing the dynamics of the correlation functions in the presence and absence of a lattice, and discuss the regimes of validity of our results. In Secs. III and IV, we focus on the dynamics of correlations in the continuum. To highlight the role of interactions, we study two types of quenches: a sudden quench from a non-interacting system to an interacting gas and the reverse quench. In the former the system evolves with a Bogoliubov dispersion while in the latter case, the dispersion is free. In Sec V. we discuss the lattice and compare and contrast the lattice from the continuum. In Sec VI., we discuss the short-range physics in the two-particle correlation functions, and relate our results to the contact. In Sec. VII, we summarize our results. Throughout this paper, we limit our discussion to sudden quenches at time $t=0$ between different initial and final states.

%In Sec. III, we calculate the dynamics of single-particle correlation function in the continuum after two types of quenches: a sudden quench from a non-interacting state, and a more general quench between two finite values of the interaction. In Sec. IV we describe the dynamics of the two-particle correlation functions both in the continuum and in a $1$D lattice. We limit our discussion to length scales comparable to or greater than the healing length of the condensate. By considering simple instructive cases, we illuminate the role of the excitation spectrum, and compare and contrast the lattice from the continuum. In Sec V.,

\section{Model} 
We start with the Hamiltonian for Bose gas interacting with a contact interaction:
\begin{equation}\label{boseham}
{\cal{H}} = \sum_{k}~\left(\epsilon_{k}- \mu\right)a^{\dagger}_{k}a_{k}  + \frac{g}{2\Omega}\sum_{p q k}a^{\dagger}_{p+q/2}a^{\dagger}_{k-q/2}a_{k}a_{p}
\end{equation}
where $g>0$ parametrizes the interactions, $\Omega$ is the volume, and $a_{k}$ is the bosonic annhilation operator. In three-dimensional free space, $g = 4\pi a \hbar^{2}/m$, where $a$ is the s-wave scattering length. In a lattice, $g = Ud^{D}$, where $U$ is the on-site interaction energy, $D$ is the dimension of space and $d$ is the lattice spacing.  In free space, the dispersion $\epsilon(k) = \frac{\hbar^{2}k^2}{2m}$, while in a lattice $\epsilon(k) = -2 J \cos(k d)$, where $J$ is the bandwidth and $d$ is the lattice spacing. In the former case, the sum is over all momenta while in the latter it is restricted to the first Brillouin zone. 

In Secs. III and IV, we study the effects of interactions on the dynamics of correlations in a $3$D continuum system. In Sec. V we highlight the differences between the lattice and continuum by considering a $1$D system in the presence and absence of a lattice potential. The latter choice is motivated by experiments. 

Working in the Heisenberg representation, we now derive the equations of motion that we study in the rest of the paper. At time $t \leq 0$, the system is assumed to be in equilibrium at zero temperature with $g = g_{i}$. At time $t>0$, the interactions are constant with $g = g_{f}$. We now present two approaches for deriving the equations for the correlation functions. 

\subsection{Time-dependent Bogoliubov approximation}

In $3$D, for weak interactions $na^{3} \ll 1$ (where $n$ is the total density),  the properties of Eq.~\ref{boseham} are well described by a Bogoliubov variational ansatz, where one sets the density of condensed atoms $n_{0} = \langle a_{k=0} \rangle^{2}$ and writes 
\begin{equation}\label{bogvar}
a_{k \neq 0}(t) = u_{k}(t)b_{k} + v^{*}_{k}(t)b^{\dagger}_{-k}
\end{equation}
where $b_{k}$ denotes the bosonic annihilation operator for the non-condensed atoms and has no time dependence and is formally treated as small. We choose $b_{k}$ operators such that $b_{k}|\psi_{0})\rangle = 0$ where $\psi_{0}$ is the initial state (with interaction $g_{i}$). We substitute Eq.~\ref{bogvar} into Eq.~\ref{boseham}, and discard all terms cubic or higher order in the $b_{k}$s. Following standard arguments we arrive at 
\begin{eqnarray}\label{bogcoheq}
u_{k}(t = 0) = \sqrt{\frac{1}{2}\left(1 + \frac{\epsilon_{k} + g_{i}n_{0}}{E^{i}_{k}}\right)} \\\nonumber 
v_{k}(t = 0) = -\sqrt{\frac{1}{2}\left(\frac{\epsilon_{k} + g_{i}n_{0}}{E^{i}_{k}} - 1\right)}
\end{eqnarray}
where $E^{i}_{k} = \sqrt{\epsilon_k(\epsilon_k + 2 g_{i}n_{0})}$.
At future times these coherence factors $u_{k}(t)$ and $v_{k}(t)$ will evolve, acquiring complex values, but will always satisfy $|u_{k}(t)|^{2} -|v_{k}(t)|^{2} =1$. We work in the regime where we can neglect the time dependence of $n_0$. 

The equations of motion for the $u_{k}$ and $v_{k}$ are obtained from the Heisenberg equations of motion for $a_{k}$. These equations are linear and can be readily integrated to give:
\begin{widetext}
\begin{eqnarray}\label{eom}
\left(\begin{array}{cc} 
u_{k}(t) \\
v_{k}(t) \end{array}\right) =\Biggl[ \cos(E^{f}_{k}t) \hat{I}  -
  i\frac{\sin(E^{f}_{k}t)}{E^{f}_{k}}\left(\begin{array}{cc} \epsilon_{k} + g_{f} n_{0} & g_{f} n_{0} \\ -g_{f} n_{0} & -(\epsilon_{k} + g_{f} n_{0}) \end{array}\right)\Biggr] 
\left(\begin{array}{cc} 
u_{k}(0) \\ 
v_{k}(0) \end{array}\right)
\end{eqnarray}
\end{widetext}
where $E^{f}_{k} = \sqrt{\epsilon(k)(\epsilon(k) + 2 g_{f}n_{0})}$ is the Bogoliubov dispersion where  $\epsilon_{k} = \hbar^{2}k^{2}/2m$.

\subsection{Expressions for Correlation functions}

Throughout this paper we are interested in the dynamics of two correlation functions: the non-condensed fraction ($n_{ex} = \sum_{\bf{k}\neq0}\langle a^{\dagger}_{k}a_{k}\rangle$),
and the \textit{equal time} density-density correlation function $g^{(2)}(\textbf{r} - \textbf{r}^{'})(t) = \sum_{\bf{q}}e^{i\bf{q}\cdotp(\bf{r - r^{'}})}\langle \rho_{\bf{q}}(t)\rho_{-\bf{q}}(t)\rangle$ where $\rho_{\bf{q}}(t) = \sum_{\bf{k}}a^{\dagger}_{\bf{k+q}}(t)a_{\bf{k}}(t)$.
The former is readily measured in time-of-flight experiments \cite{demarco}, while the latter is probed using Bragg spectroscopy \cite{miyakebragg, vale, huletbragg}, noise correlations \cite{altman} or by direct in-situ measurements \cite{trotzky, blochlightcone}. At zero initial temperature,  
\begin{equation}\label{exfrac}
n_{ex}(t) =\sum_{\bf{k}}|v_{k}(t)|^{2} 
\end{equation}

Our approximation of neglecting the time-dependence of $n_{0}$ is valid only as long as $n_{ex} \ll n$ for all times. The density correlations are

\begin{eqnarray}\label{g2}
g^{(2)}_\delta(t) = n^{2} + n\sum_{\bf{k}}e^{i \bf{k}\cdotp\delta}\Bigl(2|v_{k}(t)|^{2} + \\\nonumber  u^{*}_{k}(t)v_{k}(t) + u_{k}(t)v^{*}_{k}(t) \Bigr) + \tilde g^{(2)}_{\delta}
\end{eqnarray}
where $\delta = |\textbf{r} - \textbf{r}^{'}|$. 

The term $\tilde g^{(2)}_{\delta}$ is quartic in the $u_{k}$s and $v_{k}$s and arises from correlations between the non-condensed atoms. These correlations become important only at extremely short distances $\delta \sim a \sim 50$nm, which is roughly $20$ times smaller than the typical condensate healing length $\zeta = \hbar/\sqrt{mgn} \sim \mu$m. As we are primarily concerned with $\delta \geq \zeta$, we ignore corrections to the dynamics arising from this term, except in Section V. 

We now \textit{a posteriori} justify our assumptions by computing the cubic and quartic terms in Eq.~\ref{boseham}. After renormalizing the interactions to control an ultra-violet divergence, we find that this expectation value is small as long as $n_{ex} \ll n$. 

\subsection{Number-phase formulation} 

The time-dependent Bogoliubov approximation is predicated upon on the depletion being small $n_{ex}\ll n$.  This inequality is dramatically violated in $1D$, where $n_{ex}=n$.   
Following Shevshenko \cite{shevshenko}, we reformulate the time dependent Bogoliubov approximation in terms of number and phase variables formally writing $\psi(r) = \sum_k e^{ik\cdot r} a_k =  e^{i\phi} \sqrt{\rho}$.  To the extent that the hermitian operators $\rho$ and $\phi$ are well defined, they obey the commutation relations $[\rho(r),\rho(r^\prime)]=[\phi(r),\phi(r^\prime)]=0$ and
$[\rho(r),\phi(r^\prime)]=i\delta(r-r^\prime)$.  The breakdown of the Bogoliubov theory in 1D can then be understood as a consequence of the fact  that the phase fluctuations between points $r$ and $r^\prime$ diverge, $\langle \left(\phi(r)-\phi(r^\prime)\right)^2\rangle\to\infty$, as $|r-r^\prime|\to\infty$.  
Despite this divergence, the gradients of $\phi$ remain small if the interactions are weak.  

The analog of the Bogoliubov approximation thus becomes an expansion in $\nabla \phi$ and the deviation $\delta \rho=\rho-n$, where $n$ is the c-number density.  Formally, we introduce Bogoliubov operators via
\begin{eqnarray}\label{densbog}
\frac{\delta\rho_k}{\sqrt{n}}&=&(u_k+v_k)b_k+(u_k^*+v_k^*) b_{-k}^\dagger\\\label{phasebog}
2i\sqrt{n} \phi_k &=& (u_k-v_k) b_k - (u_k^*-v_k^*) b_{-k}^\dagger.
\end{eqnarray}
As we will see below, the equations of motion for $u_k$ and $v_k$ will again yield Eq.~\ref{eom}, but with $n_0$ replaced by $n$.  By construction, the density-density correlators are again given by Eq.~\ref{g2}, but now the quartic term is formally zero.  Thus Eq.~\ref{densbog} does not capture the ultraviolet structure investigated in section V.  

Before proceeding to deriving the equations of motion, it is useful to explore in a little more detail how Eqs.~\ref{densbog} and \ref{phasebog} resolve the divergences of Bogoliubov theory in 1D.  Taking the equilibrium coherence factors in Eq.~\ref{bogcoheq}, one sees that as $k\to0$, the two relevant combinations scale as $u_k+v_k\sim k^{1/2}$ and $u_k-v_k\sim k^{-1/2}$.  Thus the amplitude of the density fluctuations vanish as $k\to0$, but the amplitude of the 
phase fluctuations diverges.  In 3D the phase space for these fluctuations is sufficiently small that this divergence is unimportant.  In 1D, however, they eliminate all long range phase order.  Phase gradients (governed by $k \phi_k$), are well behaved as $k\to0$.  This latter feature will be essential for deriving the formalism.  

Finally one can calculate the the condensate fraction by looking at the long distance properties of the single particle density matrix,
\begin{equation}
g_1(r,r^\prime)=\langle \psi^\dagger(r) \psi(r^\prime)\rangle=\langle \sqrt{\rho(r)}e^{i(\phi(r^\prime)-\phi(r))}\sqrt{\rho(r^\prime)}\rangle,
\end{equation}
which can be expressed in terms of Bogoliubov operators using Eqs.~\ref{densbog} and \ref{phasebog}.  In 3D one can expand the resulting expression to quadratic order in the $b$'s, recovering $\lim_{|r-r^\prime|\to\infty} g_1^{(3D)}(r,r^\prime)\to n-\sum_k |v_k|^2$.  In 1D the phase fluctuations diverge, and $g_1^{(1D)}(r,r^\prime)\to 0$ as $|r-r^\prime|\to\infty$.

Having established that the number-phase representation has the right structure to generalize the Bogoliubov results, we now sketch the formal derivation of Eq.~\ref{eom}. Specializing to the free-space case, the Hamiltonian is
\begin{equation}\label{llham}
H=\int \!\!dr\, \frac{\sqrt{\rho} |\nabla \phi|^2 \sqrt{\rho}}{2m}+\frac{|\nabla \rho|^2}{8m\rho} -\mu \rho+\frac{g}{2}\rho^2.
\end{equation}
We substitute Eqs.~\ref{densbog} and \ref{phasebog}, and take the $b_k$'s to be small.  Truncating to quadratic order in these operators gives exactly the traditional Bogoliubov expression, but with $n_0$ replaced by $n$.  Our derivation of Eq.~\ref{eom} then goes through as before. In the long wave-length limit, one can neglect the $|\nabla\rho|^2$ term, and the truncated version of Eq.~\ref{llham} becomes the standard Luttinger liquid Hamiltonian.

As with Sec. II B, we justify our approximations by asking that the expectation values of the neglected terms are small.  Again, one focusses on the infrared behavior, as ultraviolet divergences can be controlled by renormalizing the interaction.  As one could deduce from dimensional analysis, the relevant dimensionless parameter controlling the expansion is $\gamma=(mg/\hbar^2) n^{1-2/D}$.  Thus in 1D our approximation works best at high density, while in 3D it works best at low density \cite{petrov}.

%The first term in Eq.~\ref{g2} is roughly equal to $n^{2}$ where $n = n_{0} + n_{ex}$ is the total density. The second term denotes correlations between the condensed and non-condensed atoms. In a non-interacting gas at zero temperature, $u_{k} = 1$ and $v_{k} = 0$, therefore $n_{ex} = 0$ and $g^{(2)}_\delta = n^{2}_{0}$, independent of $\delta$. 

\subsection{Regimes of validity}

The static properties of the two-particle correlation function at zero and finite temperature have been extensively studied by Naraschewski and Glauber \cite{glauber}. Here we limit ourselves to quenches at $T=0$, where the only length scale in the problem is the coherence length of the condensate $\zeta$. Finite temperature introduces another length scale namely the thermal deBroglie wave-length $\lambda_{th} = \hbar/\sqrt{mk_{B}T}$. At very low temperatures, the mean separation between particles is on the order of $\zeta$, and the physics is interaction dominated. At high temperatures, the mean separation between the particles is set by $\lambda_{th}$ and interactions play a minor role in the properties of the gas. We expect our results to be valid as long as $T \ll \mu$. The dynamics of correlation functions at finite temperatures is an important direction for further study.  

We also emphasize at the outset that the Bogoliubov approximation \textit{collisionless}, and as such only capable of describing the initial stages of dynamics of an interacting gas. Although correlations approach steady state values within this framework at long times, the final state need not be thermal. A more sophisticated theory of equilibration would take into account ``collisions" between the Bogoliubov quasi-particles (Landau damping), which is beyond the scope of this work \cite{zng}.

%Our results should be valid as long as the temperature $T \ll gn_{0}$, or $\zeta \ll \lambda_{th}$. Once the temperature becomes large enough that 
\section{Non-condensed fraction in the $3$D continuum} 

Prior to discussing the density-density correlation function, it is instrucive to consider a simpler quantity: the non-condensed fraction. (This is simply related to the condensate fraction: $n_{0} = N - n_{ex}$, which is readily measured in time-of-flight experiments). The dynamics of this quantity following a quench will illustrate the timescales involved in reaching a steady state for a weakly interacting Bose gas. Perhaps not surprisingly, this timescale is determined by $\tau_{mf} = \hbar/g_{f}n$, the mean-field time in the final state after the quench.

\subsection{Quench from $g_{i}=0$ to $g_{f} >0$}
The conceptually simplest case is to consider a quench from $g = 0$ to some final $g >0$. From Eq.~\ref{eom}, and using $v_{k}(t=0)=0$ we find that $v_{k}(t) = -i gn \sin(E_{k}t)/E_{k}$, thus $n_{ex} = \sum_{\bf{k}}|v_{k}|^{2} = (gn)^{2}\sum_{\bf{k}}\sin(E_{k}t)^{2}/E^{2}_{k}$. 
At long times, the sum is dominated by small values of $k$, where the dispersion is linear $E_{k} = ck$, where $c = \sqrt{2gn/m}$ is the speed of sound. With this substitution, the sum is readily performed to yield the expression:
\begin{equation}\label{condfrac}
n_{ex}(t) = \frac{1}{4\pi}\zeta^{-3}(1 - e^{-4t/\tau_{MF}})
\end{equation}
where $\zeta = \hbar/\sqrt{mgn}$ is the healing length of the condensate and is proportional (up to a constant) to the coherence length of the cloud, and $\tau_{MF} = \hbar/gn$ is the characteristic relaxation time following the quench. Therefore the population of non-condensed particles grows linearly at short times, saturating on timescales $t \sim \tau_{MF}$. At long times, the number of excitations created is \textit{larger} than the predicted zero temperature equilibrium value $n_{ex, eq} = \zeta^{-3}/6\pi^{2}$. The number of excitations is comparable to those in an equilibrium gas with $T \sim gn$ but here the energy formally diverges. In an experiment this ultra-violet divergence is resolved by taking into account the finite ramp rate. In the lattice case, discussed in Section V, the finite bandwidth also provides a cutoff. 

%Associating the energy of the final state with a temperature $T \sim gn$, the number of excitations created is somewhat smaller than the corresponding equilibrium finite temperature value of $n_{ex} \sim \zeta^{-3}/\pi^{2}$. 

Although the dynamics in this case is rather simple, the physics behind it is quite interesting. Within the collisionless Bogoliubov approximation, the number of particles at each momentum $k$ must be conserved ($\langle a^{\dagger}_{k}a_{k}\rangle$). As a result the dynamics here represents a \textit{collective} oscillation of particles in and out of the condensate with different wave-vectors $k$. The exponential decay of the condensate fraction is due to a dephasing of these collective oscillations \cite{ejmthesis}. 

\subsection{Quench from $g_{i}>0$ to $g_{f} > 0$}

Next, we consider a quench from some initial interaction strength $g_{i}$ to a final value $g_{f}$. Unlike the previous case, there are now two timescales in the problem: $\tau_{i/f} = \hbar/g_{i/f}n$. The dynamics of the non-condensed fraction depends on both these times. 

Upon evaluating Eq.~\ref{eom} we find:
\begin{equation}\label{gencondfrac}
n_{ex}(t) = \sum_{k}|v_{k}(0)|^2 - g_{f}(g_{i} - g_{f})n^{2}\frac{\epsilon(k)\sin(E^{f}_{k}t)^{2}}{(E^{f}_{k})^2 E^{i}_{k}}
\end{equation} 
The first term in this expression is the number of excitations present in the system initially. The minus sign in front of the second term indicates that if $g_{i} > g_{f}$, the number of non-condensed particles decreases over time, and vice-versa if $g_{i} < g_{f}$. Alternatively, the condensate fraction grows if $g_{i} > g_{f}$ and reduces if $g_{i} < g_{f}$.

The number of \textit{new} excitations created/destroyed is thus proportional to $g_{f}\times(g_{i} - g_{f})$. If $g_{f} = 0$, then this term is identically zero and no more excitations are created or lost. The initially quantum depleted superfluid remains depleted, unless interactions can redistribute momentum. 

%To obtain the relevant timescale governing the redistribution of momentum, we evaluate the expression for $n_{ex}$ assuming a linear dispersion. Defining $\zeta = \hbar/\sqrt{mg_{i}n}$ as the healing length of the initial state, we obtain:

%\begin{eqnarray}\label{condfrac2}
%n_{ex}(t) = \frac{1}{3\pi^{2}}\zeta^{-3}\Bigl(1 - 3\pi\tilde g_{f}(1-\tilde g_{f}) t/\sqrt{\tau_{i}\tau_{f}}\\\nonumber \Bigl[I_{0}(4t/\tau_{f}) -  
%L_{0}(4t/\tau_{f})\Bigr]\Bigr)
%\end{eqnarray}
%where $\tilde g_{f} = g_{f}/g_{i}$, and $I_{0}$ and $L_{0}$ denote the Bessel function of the second kind and the modified Struve function respectively. For small arguments, $I_{0}(x) - L_{0}(x) \approx 1 - 2x/\pi + {\cal O}(x^{2})$, while for $x \gg 1$, $I_{0}(x) - L_{0}(x) \approx 2/\pi x + {\cal O}(x^{-3})$.

%Unlike the case of a quench from zero interactions, where $\tau_{MF}$ was the only relevant timescale, the dynamics here are governed by \textit{two} timescales: at short times, the number of excitations decreases linearly in time with a timescale given by the geometric mean of the initial and final interaction strengths. At long times $t > \tau_{f}$, the number of excitations created/destroyed saturates to a constant. Notice also that unlike the case for a quench from zero interactions, the excited fraction saturates \textit{algebraically} rather than exponentially. 

\begin{figure}
\begin{picture}(100, 120)
\put(-60, -5){\includegraphics[scale=0.5]{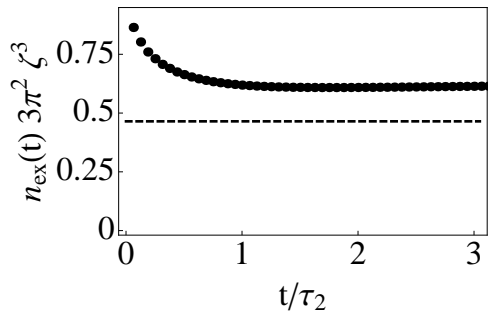}}
\end{picture}

\caption{\label{excitations} \textbf{Quasi-momentum redistribution in $3$D following a sudden change in interactions:} Plotted is the fraction of non-condensed atoms normalized to the initial excitation fraction following a sudden quench to weaker interactions $g_{f}/g_{i} = 0.6$. The excitation density is expressed in terms of $\zeta$, the healing length of the initial condensate. %Note that the analytic formula Eq.~\ref{condfrac2} (solid curve) captures the dynamics very well, particularly the long time limit. 
The characteristic relaxation time is set by $\tau_{f} = \hbar/g_{f}n$, the mean-field time of the final state. The dashed curve is the excitation fraction at the final interaction strength in equilibrium.}
\end{figure}

In Fig.~\ref{excitations} we plot the excited fraction by numerically integrating Eq.~\ref{gencondfrac} assuming the full Bogoliubov dispersion. At long times $t > \tau_{f}$, the number of excitations created/destroyed saturates to a constant value larger than the equilibrium zero temperature value at the final interaction strength $g_f$.

%We find that Eq.~\ref{condfrac2} indeed captures all the relevant features seen in the dynamics. 

%Again the number of excitations present at long times after the quench is always greater than the number of excitations in the equilibrium final state at zero temperature. Given that the Bogoliubov approximation is collisionless, there is no reason to expect the 

To summarize we see that the long time dynamics of the condensate fraction following a quench is governed by the mean-field time of the final state. At short times however there is considerable redistribution of particles into/out of the condensate due to interactions, on a timescale that depends on the initial and final interaction strength. In all cases, the final state has more excitations than the corresponding zero temperature equilibrium state. The deviations from the equilibrium zero temperature state for weak interactions are small, typically on the order of ${\cal{O}}(na^{3})$.

\section{Two-particle correlations in the continuum} 

We now turn to the dynamics of the two-particle correlation function, which will be our focus for the rest of this paper. In this section, we study two types of quenches in the continuum: a quench from zero interactions to an interacting gas and vice versa.  Our goal is to highlight the role of the Bogoliubov dispersion in determining the features observed in the density-density correlation function. 

\subsection{Short time dynamics: Diffusive to ballistic crossover} 

Consider the evolution of $g^{(2)}_{\delta}$ following a sudden quench from $g = 0$ to $g = g_{f} > 0$. At $t = 0$, $u_{k} = 1$ and $v_{k} = 0$, and  neglecting the quartic term $\tilde g^{(2)}_{\delta}$, Eq.~\ref{g2} simplifies to $g^{(2)}_{\delta}(t) = n^{2} - 2gn^{2}\sum_{\bf{k}}e^{i \bf{k}\cdotp\delta}\frac{\sin^{2}(E_{k}t)}{E^{2}_{k}}\epsilon_{k}$. Focussing on the longer range physics, we introduce dimensionless variables $\tilde\delta = \delta/\zeta$ and $\tilde t = t/\tau_{MF}$ and write
\begin{eqnarray}\label{ddcorreq}
g^{(2)}_{\delta} - n^{2} = -\frac{2}{\pi^{2}\tilde\delta}n\zeta^{-3}\int^{\infty}_{0}d\tilde k \tilde k\frac{\sin(\sqrt{2}\tilde k\delta)}{\tilde k^2 + 2} \times\\\nonumber\sin^{2}\left(\sqrt{\tilde k^2(\tilde k^2+2)}\tilde t\right)
\end{eqnarray}
where $k = \sqrt{2}\tilde k/\zeta$, and we have neglected $\tilde g^{(2)}$.  The minus sign in front of Eq.~\ref{ddcorreq} reflects the fact that for repulsive interactions the probability of finding a particle a distance $\delta$ apart is smaller than for a non-interacting system.  

\begin{figure}
\begin{picture}(100, 120)
\put(-75, -5){\includegraphics[scale=0.23]{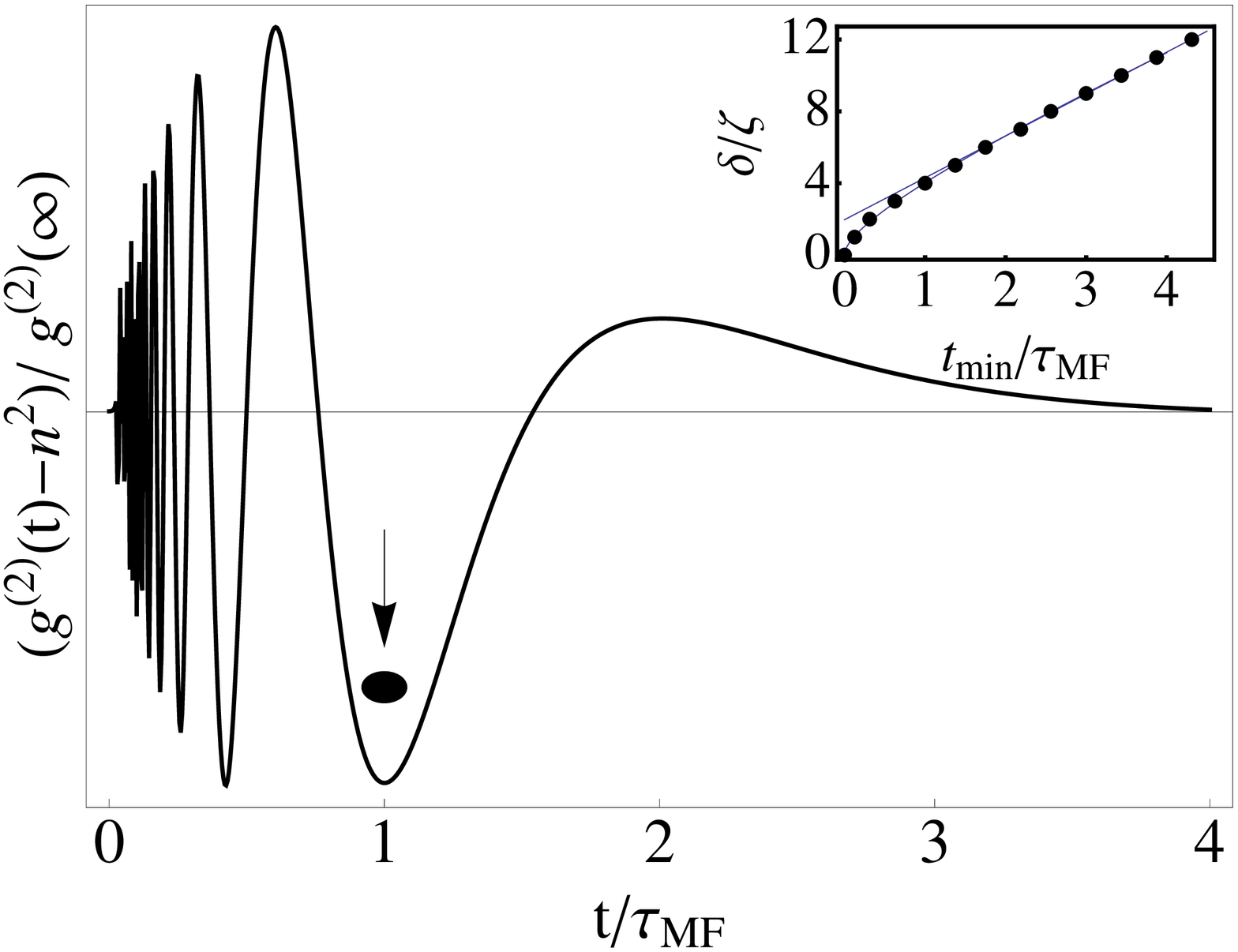}}
\put(77, 12){\includegraphics[scale=0.27]{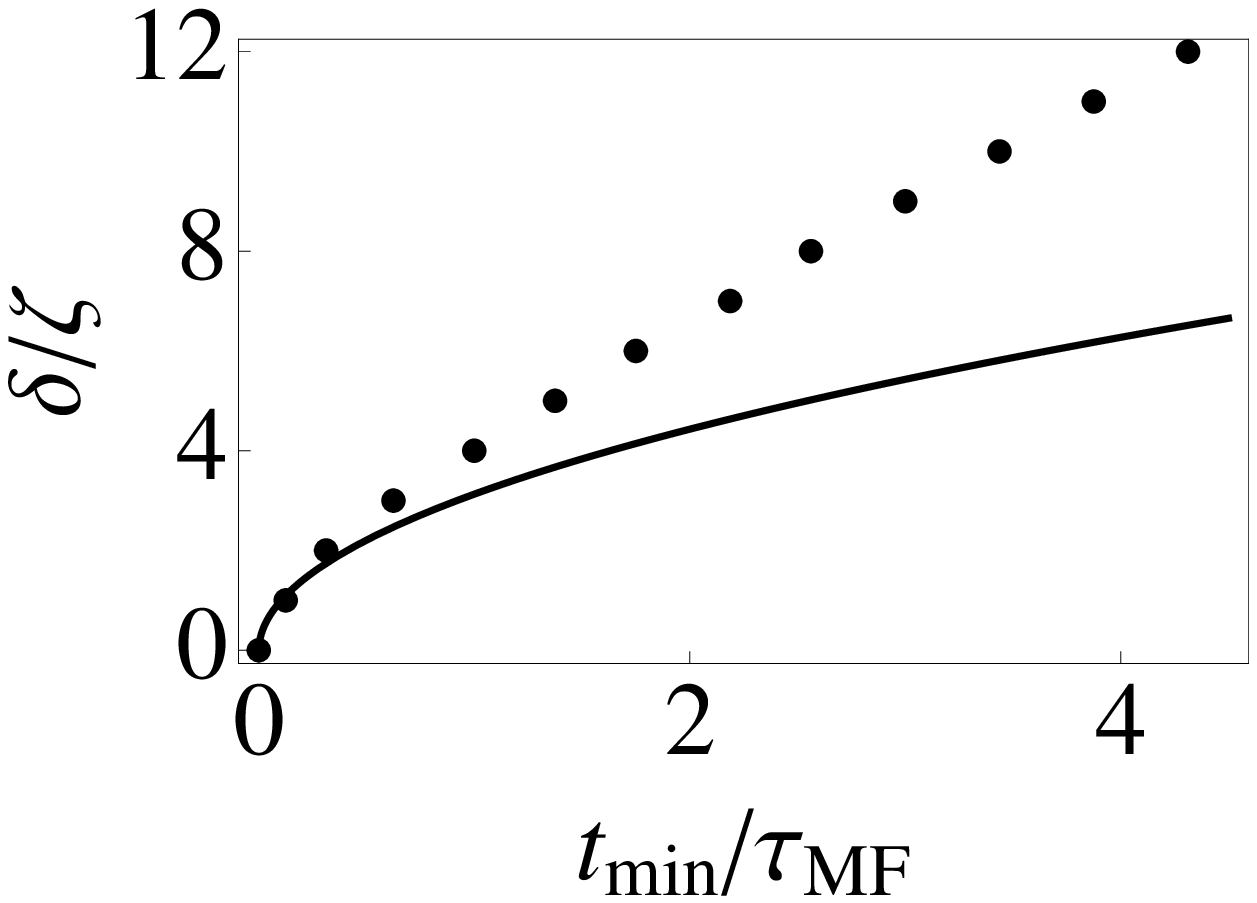}}

\end{picture}

\caption{\label{ddcorr} \textbf{Short-time dynamics of density-density correlations in $3$D} Left: density-density correlations $g^{(2)}(\delta/\zeta = 4)  - n^{2}$ normalized to the asymptotic value at long times for a quench from a non-interacting state to some final $g_{f} > 0$. Length and time is measured in terms of the condensate healing length ($\zeta = \hbar/\sqrt{mg_{f}n}$) and mean-field time $\tau_{MF} = \hbar/(g_{f}n)$ in the final state. Correlations develop in an oscillatory manner and rapidly saturate at times $t - \delta/c > \tau_{MF}$, where $\delta$ is the separation between points $\delta = |\textbf{r}-\textbf{r}^{'}|$. Inset shows the temporal location of the maximum in the correlation function (arrow on left graph) (abscissa) plotted versus $\delta/\zeta$ (ordinate). Long range correlations $\delta/\zeta \gg 1$ spread ballistically while short range correlations $\delta/\zeta \ll 1$ spread diffusively. The lines show purely linear and linear plus square root fit to the data. The slope of the linear part matches the sound velocity of the gas.  Right: Dots are the same as the inset on left. Solid line shows the location of the last maximum in the correlation function for a quench from finite interactions to $g=0$. The dynamics are purely diffusive in this case, with a diffusion constant $\approx \pi$.}
\end{figure}

In Fig.~\ref{ddcorr} we plot the evolution of the density-density correlation function $g^{(2)} - n^{2}$ as a function of $\tilde t$ for $\tilde\delta = 4$.  Focussing on the short time dynamics we find that the correlations rapidly oscillate. The temporal period of oscillations increases with time. The structure is revealed in a saddle point approximation (with the dominant wave-vectors near $k \sim \delta/t$). Within this approximation, $g^{(2)}_{\delta}$ oscillates as $\sin(\delta^{2}/t + \phi)$. 

%In order to highlight the role of high momentum states in the short time dynamics, we also show the resulting density-density correlation assuming \textit{only} a linear dispersion in the $\sin^{2}$ term in Eq.~\ref{ddcorreq}. (The $k^{2}+2$ in the denominator is still retained to ensure the integral converges). 

The temporal location of the last maximum in the correlation function ($t_{max}$) is denoted by an arrow in Fig.~\ref{ddcorr}(left). In the inset, we show how this feature disperses 
with $\tilde\delta$. When $\delta \leq \zeta = \hbar/\sqrt{mg_{f}n}$, the correlations spread diffusively ($\delta \propto \sqrt{t}$), while for $\delta \geq \zeta$, correlations spread ballistically ($\delta \propto v t$). This is indicated by the linear plus square root fit shown in the inset. The slope of the linear part matches the sound velocity of the system. Therefore, the time-evolution of the density-density correlation function can be used to extract the sound velocity of an interacting Bose gas. 

Following the quench, quasi-particles are produced locally and begin to propagate with a phase velocity $v = \omega/k$. In the continuum, the dispersion is unbounded, as a result some quasi-particles propagate very rapidly, connecting distant sites, producing rapid oscillations at short times. As we show in Section $\text{V}$, this effect is absent in the lattice, where the dispersion has a maximum velocity. The peak features on intermediate times is due to slower quasi-particles, namely the phonons leading to the crossover behavior in the density-density correlation function plotted in the inset of Fig.~\ref{ddcorr}.

%Physically, this crossover behavior arises from the Bogoliubov dispersion. At short distances, the large momentum structure of the dispersion dominates $E_{k} \sim k^{2}$, causing  diffusive dynamics; longer range correlations are governed by the linear part of the dispersion and spread ballistically. 

To further illustrate the existence of this crossover, we also consider a quench in the reverse direction from $g_{i} > 0$ to $g_{f} = 0$. Note that even for a quench to the non-interacting state, density-density correlations evolve in time (though the condensate fraction remains fixed). We find that  
\begin{equation}\label{g2eq0quench}
g^{(2)}_{\delta}(t) = g^{(2)}(t=0) - 4n\sum_{k}e^{i \bf{k}\cdotp\delta}\sin(\epsilon_{k}t)^{2}u_{k}(0)v_{k}(0)
\end{equation}

Introducing dimensionless variables as before we obtain:
\begin{equation}\label{zeroquench}
g^{(2)}_{\delta}(\tilde t) - g^{(2)}(t=0) = 2\frac{n\zeta^{-3}}{\pi^2\tilde\delta}\int^{\infty}_{0} \frac{\sin(\sqrt{2}\tilde k\tilde\delta)\sin(\tilde k^2 \tilde t)^{2}}{\sqrt{\tilde k^2+2}}
\end{equation}
%The minus sign in front of the expression Eq.~\ref{ddcorreq} signifies that upon increasing the repulsive interactions, the probability of finding a particle a certain distance apart from another particle decreases. On the other hand, for a quench to zero interactions, density density correlations grow in time. 

As in the case of the quench from $g_{i}=0$, correlations show rapid oscillations at short times, eventually saturating to some aymptotic value. As shown in Fig.~\ref{ddcorr}(right), the temporal location of the last maximum in this correlation function disperses \textit{diffusively} on all length scales as the non-interacting dispersion is $E_{k} = k^{2}/2m$ for all $k$. We find $\tilde\delta = \pi \sqrt{\tilde t_{min}}$. Thus the short time dynamics of the density-density correlation function is strongly influenced by the underlying dispersion. 

%At very small distances and short times, the dynamics is governed by the higher momentum component of the dispersion which is $k^{2}$ like. This causes short range correlations to disperse diffusively in time.  

\subsection{Long time dynamics: Exponential versus algebraic decay}

Equally interesting is the manner in which the correlations decay once they have developed. We revisit the two quenches considered in the previous subsection individually. 

For a quench from $g =0$, correlations asymptote to:
\begin{equation}\label{asymp1}
g^{(2)}_{\delta}(t \rightarrow \infty) = n^{2}  - \frac{n\zeta^{-3}}{4\pi\tilde\delta}e^{-2\tilde\delta}
\end{equation}
where $\tilde\delta = \delta/\zeta$. At short distances, $g^{(2)}_{\delta}$ appears to diverge, while at long distances, $g^{(2)}_{\delta}$ approaches $n^{2}$ from below \cite{glauber}. %On distances much shorter than $\zeta$, one must also include the contribution from $\tilde g^{(2)}_{\delta}$, which yields an even stronger divergence as $1/\delta^{2}$. Therefore, at small distances, $g^{(2)}_{\delta} \sim 1/\delta^{2} - a/\delta$ \cite{lhy}. At large $\delta$,  $g^{(2)}(\delta)$ approaches $n^{2}$ from below, as expected for a repulsive gas. 

For a quench to $g=0$, the correlations asymptote to:
\begin{equation}\label{asymp2}
g^{(2)}_{\delta}(t \rightarrow \infty) =  n^{2} + \frac{n}{6\pi^{2}\zeta^{3}\tilde\delta} \Bigl[4\tilde\delta - 3\pi( I_{2}(2\tilde\delta) - L_{2}(2\tilde\delta))\Bigr]
\end{equation} 
where $I_{2}$ and $L_{2}$ are the Bessel and Struve L functions of the second kind respectively. At short distances, $g^{(2)}_{\delta}$ approaches a constant, while at long distances, it approaches $n^{2}$ from above. 

\begin{figure}
\begin{picture}(100, 200)
\put(-30, 100){\includegraphics[scale=0.35]{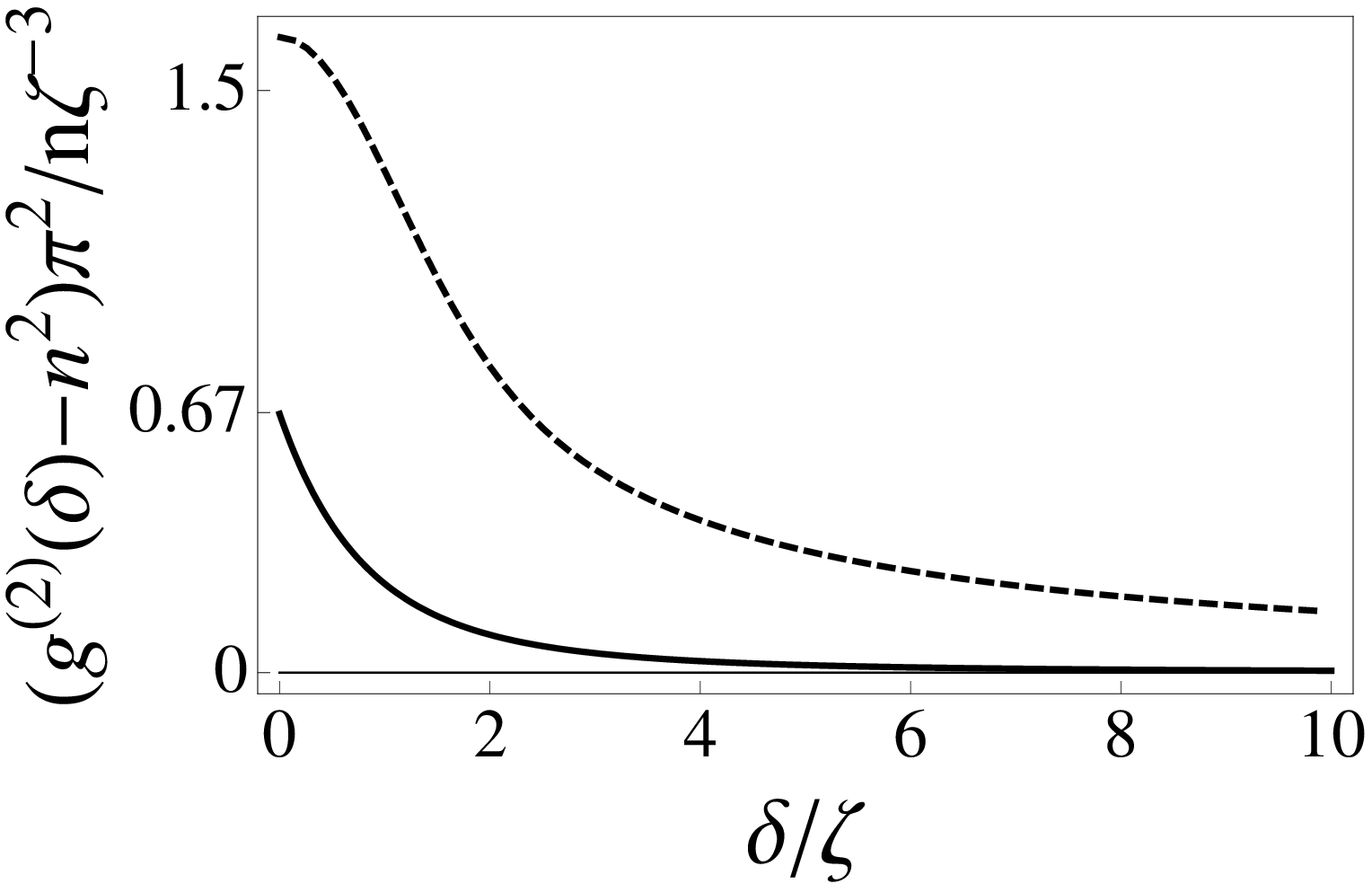}}
\put(-30, -10){\includegraphics[scale=0.35]{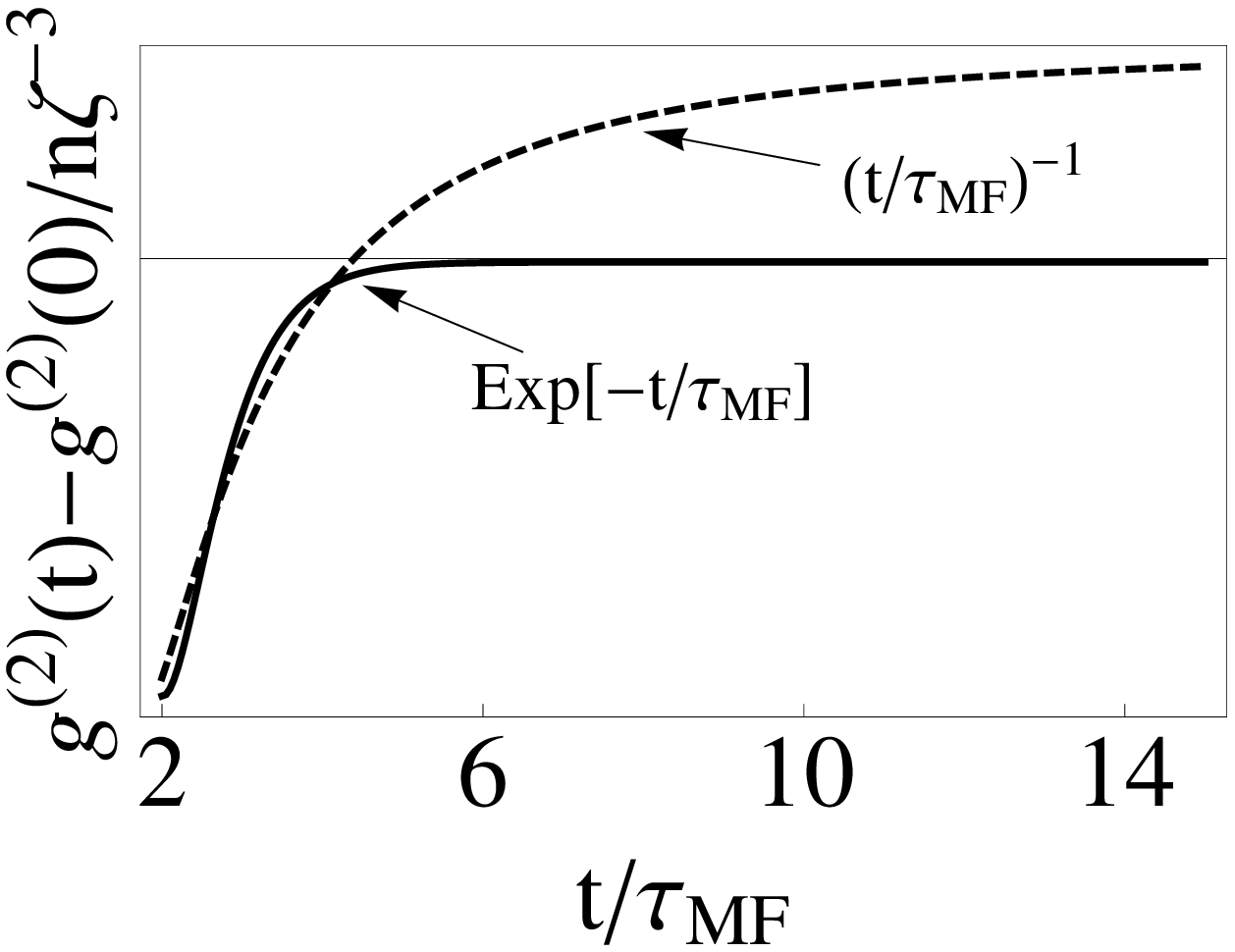}}
\end{picture}

\caption{\label{eqdens} \textbf{Long-time density-density correlation function} Top: Density-density correlation function at long times following a quench from finite interaction strength to $g=0$. The equilibrium value of $g^{2}(\delta)$ for a non-interacting condensate is $n^{2}_{0}$. The corresponding finite temperature ($T = g_{i}n$) equilibrium correlation function is shown in the dashed curve. At long distances $g^{(2)}_{\delta \gg \zeta}(t = \infty) \rightarrow n^{2} + {\cal{O}}(\tilde\delta^{-2})$ from above, whereas for short distances $g^{(2)}_{\delta \ll \zeta}(t = \infty) \rightarrow n^{2} + 4n n_{ex}$. Bottom: Long time behavior of the density-density correlation for the quenches considered in Sec.~IVA: dashed curve: quench \textit{to} zero interactions from some initial interaction strength $g>0$. solid curve: quench to a finite interaction $g>0$ from $g=0$. In either case, lengths and times are normalize to the coherence length and mean-field time in the interacting initial/final state.}
\end{figure}

In either case, the correlations \textit{do not} saturate to their equilibrium zero temperature values. In Fig.~\ref{eqdens} we compare the density-density correlation function following an interaction quench to $g=0$ with the corresponding equilibrium zero and finite temperature correlation function, assuming a temperature $T = gn$. Note that the long time density-density correlation function departs qualitiatively from the equilibrium finite temperature value.

We now discuss how the correlation functions \textit{approach} the asymptotic values above in time, finding dramatic differences depending on the low energy dispersion of the gas. 

First, we consider a quench from $g_{i}=0$ to $g_{f}=g$. Assuming that at long times, the properties of the correlation functions are governed by their low energy dispersions, we substitute $E_{k} = ck$ and solve Eq.~\ref{ddcorreq} to get:
\begin{equation}\label{g2longtime}
g^{(2)}_{\delta}(\tilde t) - g^{(2)}_{\delta}(\tilde t \rightarrow \infty) = -\frac{1}{4\pi\tilde\delta}n\zeta^{-3}e^{-2(2\tilde t - \tilde\delta)}
\end{equation}

Thus for a quench to an interacting gas, correlations decay \textit{exponentially} on times $\tilde t > \tilde\delta/2$. The timescale for this decay is set by $\tau_{MF}$ in the interacting state. 

Next we consider the reverse quench, namely from $g_{i} = g$ to $g_{f} = 0$. Performing the integration numerically, we plot the result in Fig.~\ref{eqdens}. As shown in the figure, correlations decay \textit{algebraically} as $(t/\tau_{MF})^{-1}$. 

The underlying dispersion dramatically influences how correlations decay after the quench. For the case of a quench from zero interactions, the mean-field time of the final state sets the characteristic relaxation time for the correlation function. On the other hand, for a quench to zero interactions, the mean-field relaxation time diverges ($\tau_{MF} \propto 1/\sqrt{g} \rightarrow \infty$ as $g \rightarrow 0$), producing a qualitatively different behavior in the long time dynamics of the correlation function. 

The slow decay of the correlation function for the quench to the non-interacting gas is due to the fact that the group velocity of the quasi-particles $v = \partial \omega/\partial k$ is proportional to $k$. As a result, the wave-packet spreads in space as it moves. On the other hand, for a purely linear dispersion, the wave packet retains its shape over time. The spreading leads to a much slower decay of the interference signal between the two wave-packets producing the long tail in the density-density correlation function. 

%For small $\tilde\delta$ the term in square brackets behaves linearly in $\delta$, while at large distances is scales as $1/\tilde\delta$.  Therefore, at very short distances $\delta \ll \zeta$, $g^{(2)}$ approaches a \textit{constant} value of $n^{2} + 2n/(3\pi^{2}\zeta^{3}) = n^{2} + 4nn_{ex}$, which is the Hartree-Fock result for the density correlations in a non-interacting Bose gas at finite temperatures \cite{pethick, glauber}. (Note that there is no short distance divergence in the density-density correlation function for a non-interacting Bose gas at finite temperature -- this is purely a feature of the singular repulsive interactions.) At long distances it decays to $n^{2}$ as $1/\tilde\delta^{2}$. The long time density-density correlation function is plotted in Fig.~\ref{eqdens}. 

It is tempting to connect the behavior found above and that observed numerically and experimentally in strongly interacting Bose gases \cite{blochlightcone, kollathferm, natuexp}. Analytic calculations in $1$D lattice Bose gas have shown that for sudden quenches from a Mott insulating initial state to a non-interacting final state, density-density correlations decay \textit{algebraically} as $(t/J)^{-1}$, where the hopping $J$ is the natural energy scale in the problem \cite{natuexp, kollathferm}, while experimental and numerical work on quenches in the strongly interacting regime find a much more rapid decay of correlations \cite{blochlightcone, kollathferm}. Although we do not directly model the Munich experiments here, we find very similar behavior in the weakly interacting case. 

%Curiously .. make connection with blohclightcone here!
%At very long times, only particles with $k \approx 0$ contribute to the dynamics and the correlations rapidly saturate. As in the case of the one-body correlations, the requirement for correlations to saturate is $(t - \tilde\delta/c) \gg \tau_{MF}$. 

%Finally note that density-density correlations \textit{do not} attain their zero temperature equilibrium values at long times. For the quench from $g_{i}$ to $g_{f} = 0$, we find that at long distances $g^{(2)}_{\delta \gg \zeta}(t = \infty) \rightarrow n^{2}_{0} + 2n_{0}n_{ex}$ whereas for short distances $g^{(2)}_{\delta \ll \zeta}(t = \infty) \rightarrow n^{2}_{0} + 4n_{0}n_{ex}$. Interestingly the density-density correlations at long times following a quench to a non-interacting final state do not exhibit the $1/\delta$ divergence at small distances. In Fig.~\ref{eqdens} we plot the density-density correlations at $t \rightarrow \infty$ after a quench to $g_{f} = 0$. 

%At long times, the phase of the $\sin(E_{k}t)^{2}$ oscillates rapidly and we can simply replace $\sin(E_{k}t)^{2} \rightarrow \langle \sin(E_{k}t)^{2} \rangle = 1/2$. At long times, 

\section{Lattice vs. continuum}  We now ask whether additional features are found in the presence of an underlying lattice. The conceptually simplest case to consider is a quench from finite interactions to $g=0$, in the presence and absence of a lattice. To relate our calculations to experiments, we choose a $1$D lattice with band structure $\epsilon(k) = 4J \sin^{2}(k d/2)$, but the physics described here is qualitatively similar in higher dimensions. 

\begin{figure}
\begin{picture}(100, 170)
\put(35, 80){\includegraphics[scale=0.33]{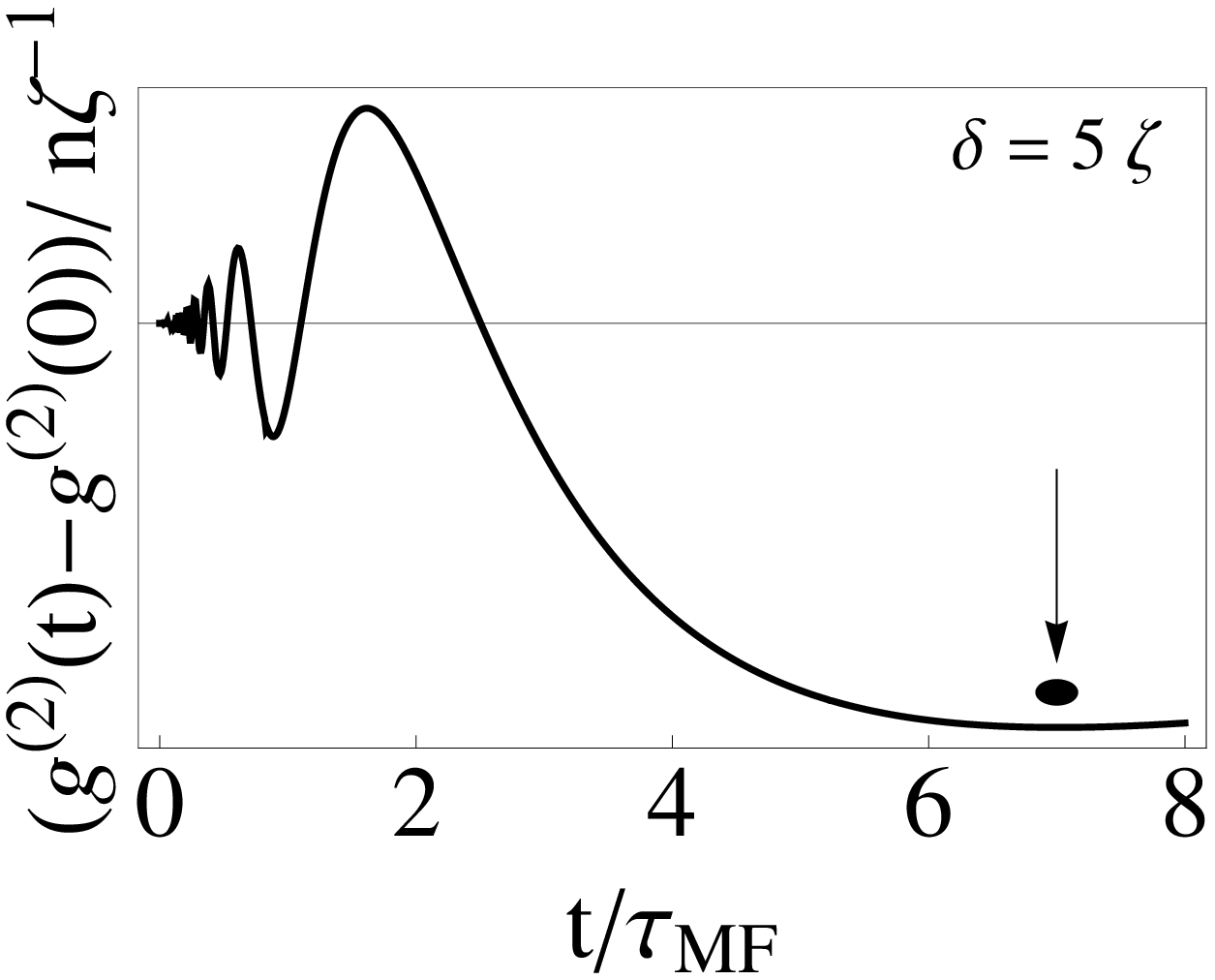}}
\put(-70, -5){\includegraphics[scale=0.4]{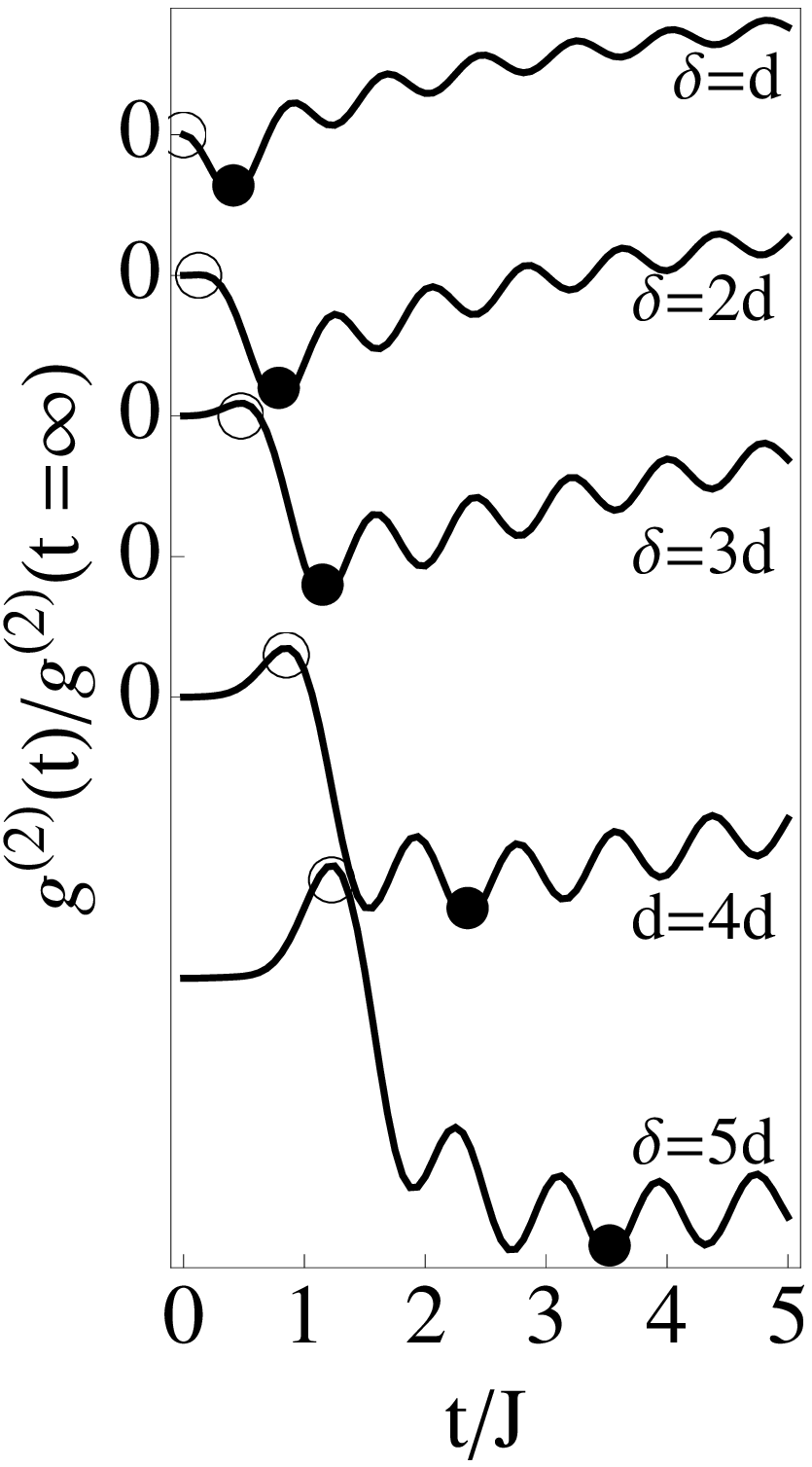}}
\put(35, -5){\includegraphics[scale=0.33]{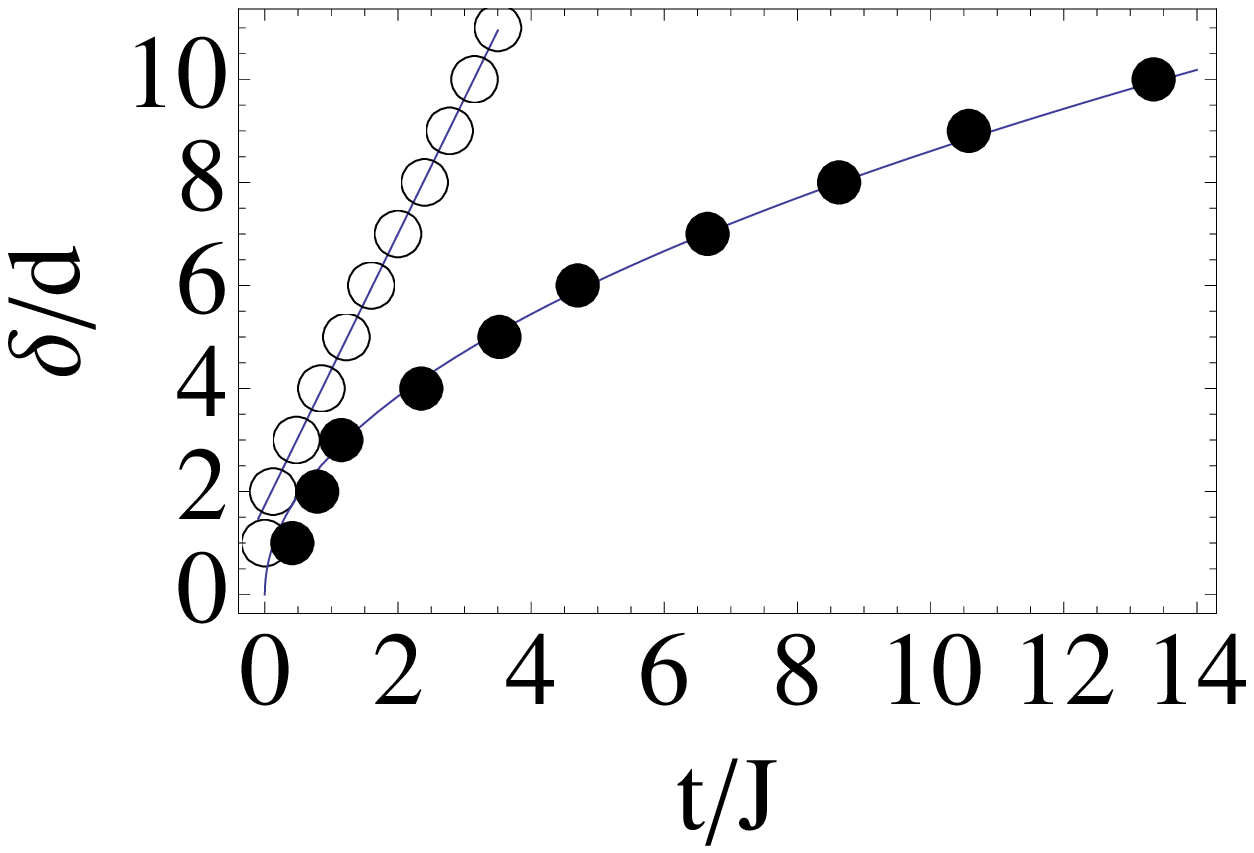}}

\end{picture}

\caption{\label{ddcorrlat} \textbf{Dynamics of density-density correlations in a $1$D lattice} (Left): Typical structure of density-density correlations $g^{(2)}(\delta)(t) - g^{(2)}(\delta)(0) $  for different values of $\delta$ normalized to the asymptotic value at long times for a quench from $g_{i} > 0$ to $g_{f} = 0$. Lengths and times are measured in terms of the lattice spacing and inverse hopping $J^{-1}$. Correlations in a lattice oscillate with a period proportional to $1/J$ and decay with a time constant independent of $\delta$. The temporal location of the first maximum ($t_{max}$) and the minimum ($t_{min}$) are indicated by open and filled circles. (Right, Top): A plot of the evolution of the density-density correlations in the $1$D continuum for the same quench at $\delta = 5 \zeta$ for comparison. The minimum in the correlation function (filled circle) disperses diffusively 
(Right, Bottom): $t_{max}$ disperses ballistically at all distances, while $t_{min}$ disperses diffusively. This latter features is the analog of the quantity highlighted on the top right.}
\end{figure}

We start by numerically integrating Eq.~\ref{g2eq0quench} in $1$ dimension, and plot the resulting density-density correlation function in Fig.~\ref{ddcorrlat}. As in the $3$D case, we find a single feature which spreads diffusively for all $\delta$. Next we substitute a lattice dispersion in Eq.~\ref{g2eq0quench} to obtain:

\begin{eqnarray}\label{g2lat}
g^{(2)}_{\tilde\delta}(t) =  g^{(2)}_{\tilde\delta}(0) + \hspace{50mm} \\\nonumber \frac{2g_{i}n^2}{\pi d}\int^{\pi}_{-\pi}dk e^{i k \tilde\delta}\frac{\sin(4J \sin^{2}(k/2)t)^{2}}{|\sin(k/2)|\sqrt{4J\sin^{2}(k/2) + 2g_{i}n}}
\end{eqnarray}
where $\tilde\delta = \delta/d$. Normalizing the energy and time units by $J$, we numerically integrate Eq.~\ref{g2lat} and plot in  Fig. \ref{ddcorrlat} the typical post-quench dynamics of the density-density correlations as a function of time, following such a quench. We choose $g_{i}n/J = 1$, none of the qualitative features discussed here are affected by this choice. 

%As the Bogoliubov theory is plagued with infra-red divergences in low dimensions (associated with the absence of long range order), we introduce a cutoff to render the integrals convergent. We find that provided the cutoff is small enough, it does not affect the \textit{qualitative} features discussed here, but merely shifts the magnitude of $g^{(2)}_{\delta}$. We pick a hard cutoff of $k_{c} = 0.00025/d$.

At long times, the correlations in the lattice and continuum both decay to their asymptotic values algebraically, but in addition, the lattice introduces  periodic oscillations. The period of these oscillations is largely independent of $\delta$, and is proportional is set by the band-width. These oscillations are a purely lattice effect and were also observed experimentally for quenches within the Mott insulating phase \cite{blochlightcone}. 

The differences between the lattice and continuum are even more striking at short times, as these reveal the high momentum structure of the underlying excitation spectrum. In Fig.~\ref{ddcorrlat} we plot the density-density correlation function for different values of $\delta$. For comparison we plot the density-density correlations in the $1$D continuum for the same quench. Note first that while the density-density correlations in the continuum build up almost immediately, there is no corresponding feature in the lattice. This is due to the fact that the lattice spectrum is bounded, correlations between two points in space take a finite amount of time to develop.

Additionally, in Fig.~\ref{ddcorrlat}, we highlight two features with open and closed circles. The open circles denote the position of the first maximum in the correlation function while the filled circles denote the minima in the correlation function. On the right we show that these features disperse differently, the position of the maximum disperses ballistically, while the position of the minimum disperses diffusively. 

The linearly dispersing maximum has no analog in the continuum and emerges purely due to the lattice band-structure, which imposes a maximum velocity ($4Jd$ \cite{cardy} in $1$D) on the spreading of correlations. The analog of this feature in the strongly interacting case has been studied in detail recently both theoretically and experimentally \cite{blochlightcone, kollathferm}. On the other hand, the diffusively dispersing minimum is also found in the continuum as shown in Fig.~\ref{ddcorr}.

For quenches to an interacting final state $g_{f} \neq 0$, the dynamics become more complicated. The minimum in the correlation functions disperses diffusively, then ballistically, similar to the $3$D continuum case. The maximum in the correlation function disperses linearly with a velocity set by the band structure for small values of $g_{f}$ crossing over to the superfluid velocity for larger $g_{f}$.

\section{Short distance structure of two-particle correlations} 

\subsection{Equilibrium properties}

Thus far we have neglected the contribution to the correlation functions arising purely from the non-condensed atoms. For $\delta \sim \zeta$, and weak interactions, these terms are of order ${\cal{O}}(a/\zeta)^2$, (for typical densities  $a \sim 50$nm  and $\zeta \sim \mu$m) and can be ignored. However for $\delta \ll \zeta$, these terms give rise to a $1/\delta^{2} $ divergence in the correlation function as a result of the singular nature of the relative wave-function of two particles interacting with a zero range interaction \cite{lhy}. The two-particle correlation function $g^{(2)}(\delta) \sim (1 - a/\delta)^{2}$ where the correlations between non-condensed particles dominate the short distance behavior and the interaction between the condensed and non-condensed particles leads to a sub leading $1/\delta$ correction which is opposite in sign.

\begin{figure}
\begin{picture}(100, 150)
\put(-60, -5){\includegraphics[scale=0.6]{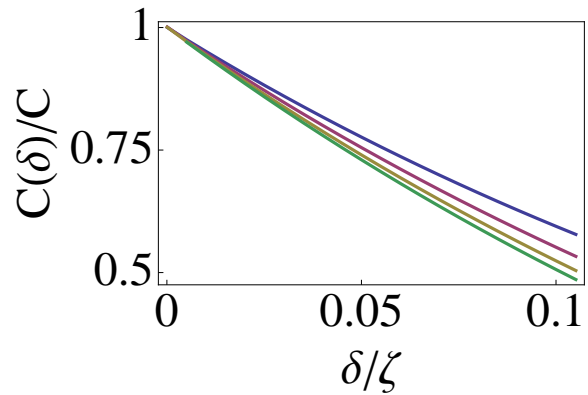}}
%\put(55, -5){\includegraphics[scale=0.33]{contactvar.eps}}
\end{picture}

\caption{\label{contact} (Color Online) \textbf{Dynamics of the contact $\cal{C}(\delta)$ in a $3$D Bose gas:} Dynamics of the position dependent contact ${\cal C}(\delta) = \delta^{2}g^{(2)}(\delta)$ (normalized to the true contact within the Bogoliubov approximation) at different times following the quench, as a function of $\delta$.  At $t <0$, the gas is non-interacting and ${\cal C}(\delta) = 0$. Immediately after the quench, the zero distance correlations respond instantaneously. Colors correspond to different times: $t/\tau_{MF} = 1$ (green), $0.5$ (yellow), $0.25$ (purple), $0.1$ (blue). We choose $a = 0.5\zeta$.}
\end{figure}

Recently Shina Tan showed that for a two-component Fermi gas interacting with a contact interaction, the short distance structure of the two-body correlation is related to the internal energy via the ``contact" defined as ${\cal{C}} = 16\pi^{2}\lim_{\delta \rightarrow 0}\delta^{2}g^{(2)}(\delta)$ \cite{tan}. For a Bose-Einstein condensate, $\cal{C}$ to leading order is $16\pi^{2}n^{2}a^{2}(1 + {\cal{O}}(\sqrt{na^{3}}))$ \cite{lhy, jincontact}. Experiments using Bragg spectroscopy to probe the Fourier transform of $g^{(2)}$ have a typical momentum resolution of $k = 2\pi/\delta_{min}$ where $\delta_{min} \sim 0.5\mu$m \cite{vale}. As a result, experiments actually probe ${\cal{C}}(\delta) = \delta^{2}g^{(2)}(\delta)$ evaluated at $\delta_{min}$.  

The correction to Eq.~\ref{g2} arising from correlations between the non-condensed atoms is:
\begin{equation}\label{g2tilde}
\tilde g^{2}_{\delta}(t) = (\sum_{k}e^{-ik\delta}|v_{k}|^{2})^{2} + |\sum_{k}e^{-ik\delta}u_{k}v^{*}_{k}|^{2}
\end{equation}

Substituting the equilibrium values of $u_{k}$ and $v_{k}$ (Eq.~\ref{bogcoheq}) into Eq.~\ref{g2tilde} we find that for small $\delta$:
\begin{equation}\label{g2tildeasymp}
\tilde g^{2}_{\delta} \sim (an)^{2}(1/\delta^{2} + (64/9-4/\pi^{2}) + {\cal}{O}(\delta^{2}))
\end{equation}
where the constant term arises from the first term on the right hand side of Eq.~\ref{g2tilde} and the divergent part arises from the second term. As $\delta \rightarrow 0$, $\lim_{\delta \rightarrow 0}\delta^{2}g^{(2)}(\delta) = a^{2}n^{2}$, or ${\cal{C}} = 16\pi^{2}a^{2}n^{2}$. Thus the Tan relations are consistent with Bogoliubov theory \cite{schackel}. 

\subsection{Non-equilibrium properties} 
 
Adding Eq.~\ref{g2tilde} to Eq.~\ref{g2} one sees that the dominant contributions to the density-density correlation function at short distances arise from 
\begin{eqnarray}\label{g2noneqshort}
g^{2}_{\delta}(t) = n \sum_{k}e^{-ik\delta}(2|v_{k}|^{2} + u_{k}(t)v^{*}_{k}(t) +  \\\nonumber  v_{k}(t)u^{*}_{k}(t))+ |\sum_{k}e^{-ik\delta}u_{k}(t)v^{*}_{k}(t)|^{2}
\end{eqnarray}
where the first term arises from correlations between the condensed and non-condensed atoms and gives a contribution which diverges as $1/\delta$ at short distances. The second term is the correlations from the non-condensed atoms alone and gives rise to a $1/\delta^{2}$ divergence at short distances.

We now consider a quench from a non-interacting gas to an interacting gas $g_{f} = g>0$. Substituting Eq.~\ref{eom} into the above expression assuming $u_{k}(0) =1$ and $v_{k}(0) = 0$, we find that
\begin{equation}\label{contact}
{\cal{C}}(\delta)(t)/{\cal{C}} = \frac{16}{\pi^{2}} \Bigl(I^{2}_{1}(\tilde\delta, t) - \frac{\pi}{2}\frac{\delta}{a} I_{2}(\tilde\delta, t)\Bigr)
\end{equation}
where we have normalized the time-dependent contact with the true equilibrium value, predicted by Bogoliubov theory. 
\begin{eqnarray}\label{i1}
I_{1}(\tilde\delta, t) = \int^{\infty}_{0}dk k \sin(\sqrt{2}k\delta)\Bigl(\frac{(k^{2}+1)\sin(E_{k}t)^{2}}{E^{2}_{k}} \\\nonumber + \frac{\cos(E_{k}t)\sin(E_{k}t)}{E_{k}}\Bigr)
\end{eqnarray}
and 
\begin{equation}\label{i2}
I_{2}(\tilde\delta, t) = \int^{\infty}_{0}dk k \frac{\sin(\sqrt{2}k\delta)\sin(E_{k}t)^{2}}{k^{2}+2}
\end{equation}
where $E_{k} = k\sqrt{(k^{2}+2)}$.

At long times, $I_{1}(\tilde\delta, t \rightarrow \infty) = \pi/8(1 + e^{-2\tilde\delta})$ and $I_{2}(\tilde\delta, t \rightarrow \infty) = \pi/4e^{-2\tilde\delta}$. For a sudden quench, at long times the non-equilibrium contact approaches ${\cal{C}}(\delta)(t \rightarrow \infty)/{\cal{C}}  =  1 - {\cal{O}}(\delta/a)$. Hence, the contact following the quench approaches its true equilibrium value on long times. 

Equally interesting is the dynamics of the contact. In Fig.~\ref{contact} we plot ${\cal{C}}(\delta)(t)/{\cal{C}}$ as a function of $\delta$ for different times. From Eq.~\ref{g2noneqshort} it is clear that the contribution to $g^{(2)}$ from correlations between the condensate and the non-condensate atoms dominates unless $\delta/\zeta \ll 1$ or $a \geq \zeta$. We choose $a = 0.5\zeta$. Immediately following the quench, the zero range correlations jump to their final value. However ${\cal{C}}(\delta)$ changes in time for finite $\delta$ and relaxes to a stationary value (denoted by the green curve) on times $t \sim \tau_{MF}$.  Therefore, although the true contact does not have any time-dependence, the experimentally relevant quantity ${\cal{C}}(\delta)$ changes in time following an interaction ramp.

%Finally we remark that \textit{even in equilibrium} the scattering length dependence of $C(\delta)$ is strongly influenced by $\delta$. For example, the $\delta \sim \mu$m scale studied in experiments,  one has to go to very large interactions in order to suppress the contribution to $g^{(2)}$ arising from correlations between the condensed and non-condensed atoms. We illustrate this in Fig.~\ref{contact}(right) where we plot ${\cal{C}}(\delta, a)$ as a function of $a$ for different values of $\delta$, after the system has equilibrated following the ramp. We find that for $\delta \sim \zeta$ (typically $\mu$m), ${\cal{C}}(a)$ varies linearly with $a$ at small $a$. The quadratic dependence of ${\cal{C}}(\delta)$ with $a$ for a weakly interacting BEC emerges only for small $\delta \sim$nm.

%In Fig.~\ref{contact} we plot the dynamics of  ${\cal C}(\delta)$ for different values of $\delta$ for different times. Note that the true contact ${\cal C}(0)$ suddenly jumps to it's equilibrium value in the final state. However the experimentally measured contact shows dynamics. For small values of $\delta$, the $C(\delta)$ approaches it's equilibirum value on a time $t \sim \tau_{MF}$. As the inset shows, in order to probe the true contact, the experimental resolution needs to be $\delta \leq 0.01\zeta \sim 10$nm.

\section{Conclusions} Relating the information contained in the fluctuations of an interacting system driven out of equilibrium to the underlying many body parameters is a challenging task. Here we study a simple model, a homogenous Bose gas at zero temperature, where one can calculate the one and two body correlations of the system following a sudden change in the interactions. Our study reveals that the dynamics of these correlations reveal a wealth of information about the underlying excitation spectrum of the system. 

We first consider the dynamics of single particle correlations, such as the condensate fraction or equivalently, the excited fraction of atoms following a sudden interaction ramp. For a quench from a non-interacting gas, we find that all the dynamics are governed by a single timescale, the mean-field time. For quenches between two interacting systems, we find two timescales governing the long and short term dynamics of the gas.  Moreover we find that the excited fraction at long times after the quench is always greater than the expected equilibrium value.

We then consider the dynamics of the two particle correlation function. For quenches between interacting initial and final states, we find a non-trivial crossover between diffusive spreading of short-range correlations and the ballistic spread of long range correlations. We relate this crossover to the underlying Bogoliubov dispersion. Furthermore, we show that the interacting dispersion leads to a much more rapid decay of correlations as compared to the free case. 

We then consider the case of a Bose gas in a lattice and discuss the additional features that arise in the correlations from the band structure. Comparing a $1$D lattice with the $1$D continuum case, we find that the band structure leads to an additional linearly dispersing feature in the correlation functions. For a non-interacting final state, the velocity is set by the bandwidth, while for an interacting final state, it crosses over to the sound speed of the gas. 

Finally we discuss the dynamics of the contact following a sudden quench. We find that while the true zero range contact instantaneously takes on the new equilibrium value, the finite resolution of an experiment will make the contact appear time dependent. 

\textit{Acknowledgements.---} This work was supported by a grant from the Army Research Office with funding from the DARPA OLE program and by the National Science Foundation through grant Nos. $1066293$ and PHY $1125915$. We would like to thank Kaden Hazzard, Philip Makotyn and Eric Cornell for helpful discussions. SN is grateful to the Kavli Institute of Theoretical Physics (KITP) for its hospitality and would like to thank the organizers and participants of the KITP program entitled \textit{Quantum Dynamics in Far from Equilibrium Thermally Isolated systems} for valuable discussions.

\end{document}